\documentclass{emulateapj}

\shorttitle{Detection of anomalous emission in Perseus} 
\shortauthors{Watson et al.}

\begin{document}

\title{Detection of anomalous microwave emission in the Perseus molecular
cloud \\ with the COSMOSOMAS experiment}

\author{R. A. Watson\altaffilmark{1,2}, 
R. Rebolo\altaffilmark{2}, J. A. Rubi\~no-Mart\'\i n\altaffilmark{2}, S. Hildebrandt\altaffilmark{2}, C. M.
Guti\'errez\altaffilmark{2}, S. Fern\'andez-Cerezo\altaffilmark{2}, R. J.
Hoyland\altaffilmark{2} and E. S. Battistelli\altaffilmark{2}} 

\affil{
    $^{1}$ Jodrell Bank Observatory, Macclesfield, Cheshire, SK11 9DL, UK \\
    $^{2}$ Instituto de Astrofis\'{\i}ca de Canarias, 38200 La Laguna, Tenerife,
Spain;\\
raw@iac.es, rrl@iac.es, jalberto@iac.es, srh@iac.es, cgc@iac, scerezo@iac, rjh@iac, 
elia.stefano.battistelli@roma1.infn.it}

\begin{abstract}
We present direct evidence for anomalous microwave emission in the Perseus
molecular cloud, which shows a clear rising spectrum from 11 to 17 GHz in
the data of the COSMOSOMAS experiment. By extending the frequency coverage
using {\it WMAP} maps convolved with the COSMOSOMAS scanning pattern we
reveal a peak flux density of 42 $\pm$ 4 Jy at 22 GHz integrated over an extended
area of $1.65\degr\times1.0\degr$ centered on RA=$55.4\degr\pm 0.1\degr$ and
Dec=$31.8\degr\pm 0.1\degr$ (J2000). The flux density that we measure at this
frequency is nearly an order of magnitude higher than can be explained in
terms of normal galactic emission processes (synchrotron, free-free and
thermal dust). An extended {\it IRAS} dust feature G159.6-18.5 is found near
this position and no bright unresolved source which could be an ultracompact
H\,{\sc ii} region or gigahertz peaked source could be found. An adequate
fit for the spectral density distribution can be achieved from 10 to 50 GHz 
by including a very significant contribution from electric dipole emission
from small spinning dust grains.

\end{abstract}


\section{Introduction} 

The search for `anomalous microwave emission' or `dust-correlated microwave
emission' was started by the surprising statistical correlation of {\it
COBE} DMR observations at centimetric wavelengths with DIRBE dust emission
at 140\micron~ \citep{Kogut96}. \citet{Leitch97} found a similar correlation
with the OVRO ring observations at 14.5 and 32 GHz. Ordinary thermal
emission from dust at these relatively low frequencies is expected to be
orders of magnitude lower than the intrinsic CMB fluctuations. It was
initially suggested that this anomalous emission had its origin with
free-free processes associated with the warm ionized interstellar medium.
The lack of the expected H$\alpha$ from this ionized gas would require
temperatures hotter than $10^6$K. However \citet{DraineLazarian98} argued
against such high temperature on energetic grounds and proposed an
alternative mechanism based on electric dipole emission from small dust
grains. Further evidence for anomalous microwave emission has been found in
data from other CMB experiments such as Saskatoon \citep{deOliveira97},
19GHz \citep{deOliveira98}, Tenerife \citep{deOliveira99, deOliveira02,
deOliveira04}, Python V \citep{Mukherjee03}, {\it WMAP} \citep{Lagache03,
Finkbeiner04a} and the Green Bank Galactic Plane Survey (GBGPS)
\citep{Finkbeiner04b}.

Despite this growing evidence, the exact nature and distribution of this new
foreground remains elusive and part of this is because there is no
comparable survey below 20 GHz to complement the {\it WMAP} data. For
instance, \citet{Bennett03a} find that {\it WMAP} foregrounds can be fitted
by a proposed flatter component of synchrotron. \citet{Finkbeiner04a} and
\citet{deOliveira04} point out that Tenerife and GBGPS data provides
evidence for rising spectrum at 10 and 15 GHz which is incompatible with
synchrotron emission. A similar survey to the Tenerife experiment
\citep{Gutierrez00} in frequency and sky coverage but with degree scale
resolution is now available with the COSMOSOMAS experiment \citep{Cerezo05,
Hildebrandt05} of the Instituto de Astrof\'isica de Canarias (IAC).

The COSMOSOMAS experiment
consists of two circular scanning instruments
operating at the Teide observatory (altitude 2400 m, Tenerife). They have
produced 0.8-1.1\degr~ resolution maps of $\approx$ 10000 square degrees in
the sky in four frequency bands (centered at 10.9, 12.7, 14.7, 16.3 GHz). The
average sensitivity of each map is in the range $80 - 120 \mu$K per beam.
The basic observing strategy is described in \citet{Gallegos01}. Details of
the survey by the two instruments where a significant dust correlation
is found at $|b| > 20\degr$ is given in \citet{Cerezo05, Hildebrandt05}. In
this letter we present results of a search for sites of dust-correlated
emission with a rising flux density spectrum in the first release of the COSMOSOMAS
maps.

\section{The search for areas of rising spectrum with COSMOSOMAS}

Since we are focusing on bright discrete sources we chose to generate a list
of possible sources with sextractor \citep{BertinArnouts96} using a
detection threshold of 4 sigma on the more sensitive 13 GHz COSMOSOMAS map
($+22.0\degr < $Dec$ < +45.0\degr$).  Since in the Auriga-Perseus galactic
crossing the confusion is lower and the sources are more clearly separated,
we have centered our search for sources of rising spectrum in this region
which is shown in fig~1. 

\begin{figure*}\center
\includegraphics[angle=-90]{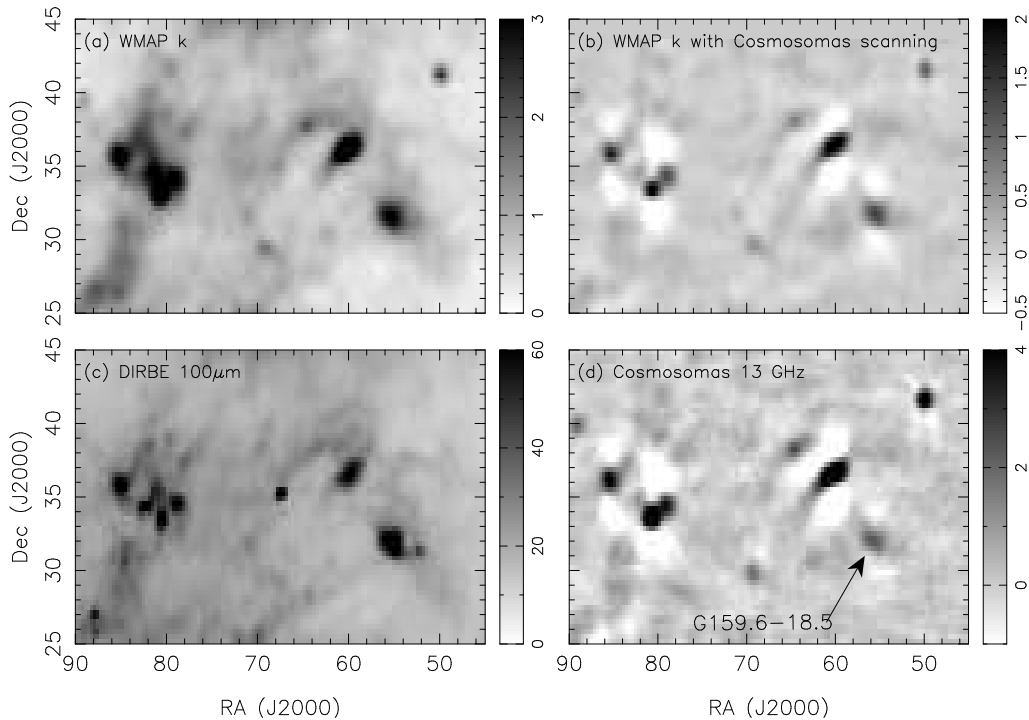}
\caption{Comparison of various survey maps over the Auriga-Perseus region.}
\end{figure*}

A few of the objects turned out to be artifacts on close inspection and 
fluxes were computed for all remaining true sources in each of the COSMOSOMAS
frequency bands using a 3 by 3 pixel box (1\degr RA by 1\degr Dec). One
source (RA = $55.4\pm 0.1\degr,$ Dec = $+31.8\pm 0.1\degr$ J2000) stood out
with a clear steep rising spectrum between 11 and 17 GHz with an index of
$\alpha = +1.4 \pm 0.3$ (S$_\nu \propto \nu^{\alpha}$). This source lies in
the Perseus OB2 area at an intermediate latitude b=-18$\degr$ and is also
seen as a bright source in the 100\micron~ DIRBE map which appears to
correspond to the extended {\it IRAS} source G159.6-18.5. It also appears to
be extended in the COSMOSOMAS maps with a FWHM $\approx 2.0\degr$. The
average temperatures of this pixel box are shown in Table~1 showing the
almost constant thermal spectrum over the COSMOSOMAS frequencies.

To make a comparison, the {\it WMAP} data \cite{Bennett03a} have been
degraded to the common COSMOSOMAS resolution (set at 1.12\degr~), convolved
with the simulated circular lock-in analysis and pixelated in a regular
1/3\degr~ Cartesian grid in RA and Dec. The lockin analysis includes the
removal of the first 7 harmonics to suppress the 1/f noise from receiver and
atmosphere. The effect on the WMAP maps is shown by comparing the original
map fig.1(a) and the convolved map fig.1(b), where the loss of broad angular
features is evident with the creation of negative lobes north and south of
bright objects in the center of the map. This figure also gives a visual
example of this dust correlation through the strikingly similar {\it WMAP} K
(22 GHz) (fig.1a) and DIRBE 100\micron (fig.1c) maps in the galactic
Auriga-Perseus region.

Two of the detected sources in this area are bright diffuse H\,{\sc ii}
regions, NGC 1499 and IC410 that we use as a calibration check. Using 1420
MHz \citep{Reich88} and the five {\it WMAP} template maps we verified that
the emission from these objects has a spectral index of $-0.15\pm .02$ and
$-0.20\pm .03$, respectively. These indices are consistent with that
expected for free-free emission of -0.12 between 11 and 17 GHz. For both
sources we find the measured COSMOSOMAS fluxes are generally lower by
$\approx$ 6\% than the predicted values from the fits. This is consistent
with the absolute calibration error of the COSMOSOMAS maps with is better
than 10\% in the Auriga-Perseus region.

In order to get the dust correlated emissivity ($\mu$K [MJy]$^{-1}$ sr) for
G159.6-18.5 and to quantify the anomalous microwave emission we have carried
out a correlation with the 100\micron~ DIRBE map fig.1(c) passed through the
COSMOSOMAS scanning. We chose a 11 by 11 pixel box, an area of
$3.1\degr\times 3.7\degr$, centered on the pixel nearest to the center of
this source. This covers the bulk of the emission allowing a good range of
pixel values without including much background. The correlation is
calculated by the linear fit of temperature ($\mu$K) of the COSMOSOMAS and
{\it WMAP} pixels to the intensity (MJy/sr) of the DIRBE pixels using
a standard least squares method. Table~1 gives the values for the slopes and
errors showing a tight correlation with DIRBE. The equivalent analysis with
the 408 and 1420 MHz templates suggests free-free emission is present but
with less significance since this emission appears more compact and displaced
with respect to the DIRBE emission.

\section{G159.6-18.5 and its spectral energy distribution}

G159.6-18.5 lies within the Perseus Molecular cloud complex at a distance of
260 pc \citep{Cernicharo85}. It appears in {\it IRAS} maps as a slightly
broken ring with a diameter of $\approx 1.5\degr$ composed of several
diffuse knots around the rim and a couple in the center, all with
intensities of 100-200 MJy/sr at 100\micron. These knots are still bright at
12\micron~ with intensities of 5-10 MJy/sr suggesting the presence of
transiently heated small dust grains. 

At first it was thought to be a supernova remnant (SNR) by
\citet{Fiedler94}, but it was later argued to be a H\,{\sc ii} region driven
by the O9.5-B0 V star HD 278942 \citep{deZeeuw99} found to be at the center
of the ring. \cite{Andersson00} have made an extensive overview of the
nature of this dust ring including interferometric observations at 408 and
1420 MHz and conclude the ring is a ruptured blister H\,{\sc ii} region
which has emerged from the outer edge of the cloud. Their flux density
estimates are $F(408 MHz) = 2.0^{+1.8}_{-0.1}$ Jy and $F(1420 MHz) =
1.4^{+4.3}_{-0.1}$ Jy allowing for incomplete sampling of the $u-v$ plane.
This gives a spectral index of $\alpha = -0.3^{+0.3}_{-1.1}$, which is
consistent with an optically thin H\,{\sc ii} region ($\alpha = -0.1$). They
also found many radio sources, but all are steep spectrum apart from the
flat spectrum quasar 4C+32.14, which appears at the NW edge of the ring.
This was removed from all the maps under consideration here assuming a flux
of 2 Jy at all frequencies.

\begin{deluxetable}{rrrr}
\tablecaption{Spectrum of G159.6-18.5\label{Spectrum}} 
\tablehead{\colhead{$\nu$} & \colhead{Temperature} &
\colhead{Correlation} & 
\colhead{Flux} \\ 
\colhead{(GHz)} & \colhead{(mK)} &
\colhead{($\mu$K MJy$^{-1}$ sr)} & \colhead{(Jy)} }

\startdata
0.408  & 1346  $\pm$ 214   & 56000 $\pm$ 70000 & 6.3 $\pm$  7.8 \\
1.420  & 144   $\pm$ 14    & 4260  $\pm$ 1410 &  7.3 $\pm$  2.0 \\
10.9   & 2.264  $\pm$ 0.096  & 59.6   $\pm$ 0.8 & 17.0 $\pm$  0.1 \\
12.7   & 1.798  $\pm$ 0.098  & 45.1   $\pm$ 1.9 & 18.7 $\pm$  0.4 \\
14.7   & 1.839  $\pm$ 0.072  & 44.5   $\pm$ 2.0 & 26.2 $\pm$  0.6 \\
16.3   & 1.682  $\pm$ 0.065  & 41.9   $\pm$ 4.0 & 32.6 $\pm$  1.5 \\
22.8   & 1.195  $\pm$ 0.035  & 33.9   $\pm$ 0.3 & 42.3 $\pm$  0.2 \\
33.0   & 0.547  $\pm$ 0.016  & 15.7   $\pm$ 0.3 & 40.3 $\pm$  0.4 \\
40.9   & 0.296  $\pm$ 0.009  &  7.7   $\pm$ 0.3 & 33.9 $\pm$  0.7 \\
61.3   & 0.134  $\pm$ 0.004  &  3.3   $\pm$ 0.3 & 34.7 $\pm$  1.8 \\
93.8   & 0.116  $\pm$ 0.003  &  2.7   $\pm$ 0.4 & 77.5 $\pm$  4.3 \\
1250.  &            ~        &        ~       &   96800 $\pm$  700. \\
2143.  &            ~        &        ~       &  131000 $\pm$  1250. \\
3000.  &            ~        &        ~       &  64400 $\pm$  100.\enddata
\end{deluxetable}

To investigate the emission mechanisms occurring in this object we need to
estimate its spectral energy distribution (SED) over a wide range of
frequencies. Since flux on large angular scales is lost due to the
COSMOSOMAS lock-in analysis we are forced to fit the data to a convolution
of a sky density flux model. We adopted a simple elliptical Gaussian for G159.6-18.5
based on a fit to the original {\it WMAP} Ka map. We obtained a FWHM ellipse
for the unconvolved model of $1.6\degr\times1.0\degr$ PA = 51\degr. The
expected beam spread function for COSMOSOMAS was then fitted to all the
template maps using a least squares method allowing for a zero level offset.
For these fits we obtained the amplitudes and subsequently the fluxes, which
are listed in table~1 and plotted in fig.2. The two low frequency surveys do
not show the extended emission, but unresolved emission from the center of
the ring so we used the sky model corresponding to a point source. The error
in the estimate of the FWHM of the sky model of 0.1\degr~ leads to an
overall absolute calibration error of 10\% although the relative calibration
between frequencies is better than 2\%.

\section{Limits on standard emission mechanisms}

The main galactic microwave emission mechanisms are synchrotron, free-free and
thermal dust emission. Synchrotron can be estimated from the low frequency
surveys, free-free from H$\alpha$ emission and thermal dust from DIRBE.

The values from the 408 and 1420 MHz surveys for G159.6-18.5 suggest no
significant synchrotron emission and some free-free emission at the these
frequencies, consistent with the flux density measurements made by
\citet{Andersson00} inside the ring. A ``bean'' shaped patch of H$\alpha$
emission inside the dust ring is detected in this region by the Wisconsin H
Alpha Mapper project (WHAM) \citep{Haffner99}. This suggests the H$\alpha$
emission is due to diffuse H\,{\sc ii} and is shining through a thinner or
ruptured part of the dust shell. Estimating free-free emission from the
H$\alpha$ emission is difficult due to the high extinction which also varys
strongly across the ring and the unknown fraction of dust in front of the
H\,{\sc ii} region, but we only require a rough indication of the
level of emission. We adapted the method that \citet{Dickinson03} describe to
estimate galactic free-free emission from H$\alpha $. Assuming an extinction
factor $Rv=4.0$ as expected for molecular clouds we use an expression for
H$\alpha$ extinction A(H$\alpha$) = 3.33 E(B-V) together with the E(B-V)
maps of \citet{Schlegel98} and find a mean corrected H$\alpha$ intensity
over the ring of 120 Rayleighs. At 22 GHz, close to the peak source
emission, the free-free brightness temperature to H$\alpha$ intensity ratio
($T^{ff}_b/I_{H\alpha}$), assuming electron temperature 8000 K, is $12.3\mu
K/$Rayleighs. Integrating over the area of the ring this corresponds to a
flux density of 4.5 Jy, similar to the 5.6 Jy estimated by extrapolating
from the 1420 MHz flux density derived from the Reich and Reich map. This is
nearly an order of magnitude below the estimated flux density of 42 Jy from
the {\it WMAP} K data.

Thermal dust emission can be fitted to the DIRBE and {\it WMAP} W points as
shown in fig.2, with a dust temperature of 19 K with opacity at 100\micron
($\tau_{100}$) of $6.0\times 10^{-3}$ and emissivity index 1.55. This is
what is expected for dust in a ``warm neutral medium'' (WNM). This WNM
region might be generated by interaction with the nearby B0 star. A very cold
dust grain model as tried by \citet{Casassus04} with CBI observations
of anomalous microwave emission from the Helix nebula would not be
consistent with COSMOSOMAS and {\it WMAP} points. 

\begin{figure*}\center
\includegraphics[angle=-90]{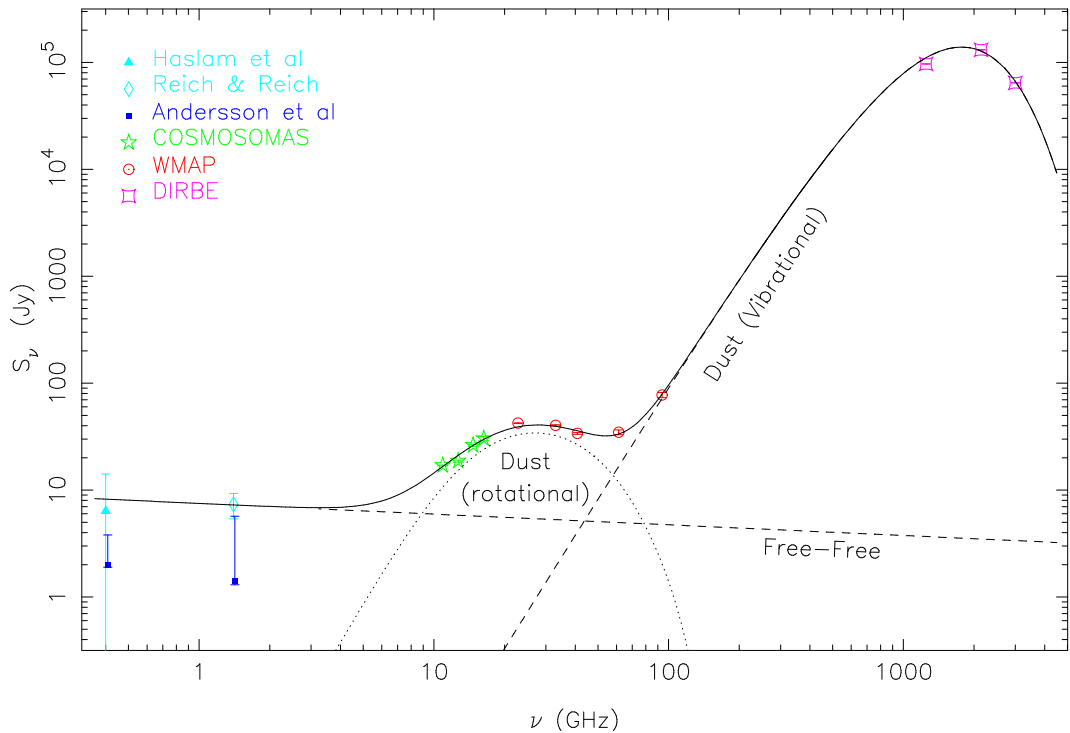} \caption{Spectrum of G159.6-18.5. Points
are shown for COSMOSOMAS, {\it WMAP}, DIRBE and low frequency (408 \& 1420
MHz) surveys convolved to give COSMOSOMAS equivalent points. The size of the
symbols correspond to the sky model uncertainty of 10\%. Estimates are made
of three foregrounds; free-free, rotational dust and vibrational dust.
Vibration dust corresponds to $T_{dust}$ = 19.0K and emissivity index 1.55.
A Toy dust model similar to the one used by Finkbeiner is used which
consists of a linear combination of Draine and Lazarian models for Warm
Neutral Medium and Molecular Cloud (0.8*WNM + 0.3*MC). }
\end{figure*}

There are two known types of object which show a convex spectrum at
microwave frequencies; ultracompact H\,{\sc ii} regions (UCH\,{\sc ii}) and
Gigahertz Peaked Spectrum (GPS) sources which we assess as possible
sources to provide the excess flux density at 22 GHz.

UCH\,{\sc ii} regions produce free-free emission which is self-absorbed at
low frequencies due to extremely high densities and the high dust opacity
can hide the H$\alpha$ emission. Indeed \cite{McCulloughChen02} suggested
that an UCH\,{\sc ii} could explain the microwave emission detected by
\cite{Finkbeiner02} in the diffuse H\,{\sc ii} region LPH 201.663+1.643, but
it did not seem to have the necessary FIR luminosity and so anomalous
emission was not ruled out.  In the case of the Perseus cloud its distance
is known and UCH\,{\sc ii} regions being amongst the brightest objects at
100\micron~ would have a flux density in excess of 1 MJy. None of the {\it IRAS}
point sources within 1\degr of the COSMOSOMAS source center have {\it IRAS}
colors which meet \cite{WoodChurchwell89} criterion for UCH\,{\sc ii}
regions and are all less than a few hundred Janskys. Such a source could be
behind the cloud at a distance of $\gtrsim 10$ kpc, but at galactic latitude
-18\degr this seems very unlikely. 

GPS sources are high red-shift radio sources in which the radio jets have
been highly confined and the synchrotron emission is self-absorbed. The
majority of GPS sources have a peak around 1 GHz, but there are a few rare
ones such as 1459.9+3337 \citep{Edge98} which peak at 30 GHz. A GPS source
having an index of +1.4 on the low frequency side as suggested by COSMOSOMAS
data would have a flux density at 4.85 GHz of $\approx$ 5 Jy. A search in
the Green Bank survey (GB6) \citep{Gregory96} within 1.5\degr of the
COSMOSOMAS source center finds the brightest source, apart from 4C+34.14,
J0340+3209 which is 0.47\degr away with a flux density of 482 mJy. A cross
check in the NVSS catalogue \cite{Condon98} over the same area finds this
source again as the brightest with a flux density of 1.6 Jy showing it to be
a steep source with no significant emission at the COSMOSOMAS frequencies.

The extended nature of the COSMOSOMAS source argues against a radio point
source and hence against either of these objects.

\section{Discussion}

\citet{DraineLazarian98, DraineLazarian99} have made detailed calculations for
the expected spectrum from electric dipole emission of very small spinning
dust grains and magnetic dipole radiation from hot ferromagnetic grains.
It is still uncertain the exact mass and dipole moment distribution
these grains will have. 

As a test of the spinning dust hypothesis we can check if it can produce the
measured spectrum with respect to the thermal dust emission. Draine and
Lazarian's predictions are made in terms of emission in intensity per
hydrogen column density in units of Jy/sr cm$^2$/H. We estimate a column
number density of hydrogen atoms N(H) = $1.3\times 10^{22}$ using the
canonical factor of $2.13\times 10^{24}$ H cm$^{-2}$ = unit $\tau_{100}$ as
in \citet{Finkbeiner04b}. There is not enough information to make a detailed
model of emission taking into account the varying conditions throughout the
cloud, so we use an integrated sum of the WNM model A of Drain and Lazarian
for the dense region facing the B0 star with a contribution of the
``molecular cloud'' (MC) model for dust warmed further in the cloud.
Assuming the solid angle of the emission is that of the sky model used in
section 2 ($1.6\degr\times1.0\degr$) we find a qualitatively good agreement
with the observations using 0.8 WNM and 0.3 MC of the predicted curves (see
figure 2). Although we have not attempted it, there maybe magnetic dipole
emission models which can fit the data also.

This anomalous signal in this cloud is so evident because the competing
free-free emission is low. If the free-free emission was five times
brighter, for instance if the star was more massive and more embedded, then
the detection would be marginal. Yet the stars interaction seems important
to excite the thicker parts of the dust shell still embedded in the cloud.
It seems unlikely that objects such as G159.6-18.5 could be at high galactic
latitudes, but it may prove to be a relatively close example of anomalous
emission which can be studied in detail. The bright FIR luminosity and
associated microwave continuum emission may mean other such objects exist at
low galactic latitudes but have been classified as UCH\,{\sc ii} regions. 

\section{Conclusion}

Our search for rising spectrum sources in the COSMOSOMAS data finds
anomalous microwave emission associated with the {\it IRAS } dust feature
G159.6-18.5 in the Perseus molecular complex. We measure a rising flux
spectrum from 11 to 17 GHz with a spectral index of +1.4. Extended frequency
coverage using cross calibrated WMAP data reveals the expected spectral roll
off at both high and low frequencies required for a solid detection of
anomalous emission. The bulk of the emission seems to come from the
thicker dusty wall on the molecular cloud side of the ring rather than the
center of the ring itself.

We discard an ultracompact H\,{\sc ii} region or a Gigahertz Peaked Spectrum
source as the cause of this microwave emission since they would appear as
point sources whereas both COSMOSOMAS and {\it WMAP} find an extended
diffuse feature. A simple combination spinning dust models can explain the excess
microwave emission.

We acknowledge the support of the Instituto de Astrof\'isca de Canarias and
the staff of Teide Observatory for the construction and operation of the 
COSMOSOMAS experiment. Partial funding was provided by grant AYA2001-1657 of
the Spanish Ministry of Science and Education.

\end{document}